\documentclass[runningheads]{llncs}

\usepackage{cite}
\usepackage{amsmath,amssymb,amsfonts}
\usepackage{algorithmic}
\usepackage{graphicx}
\usepackage{textcomp}
\usepackage{xcolor}
\usepackage{cuted}
\usepackage{tikz}
\usepackage{url}
\usepackage{subcaption}
\usepackage{wrapfig}
\usetikzlibrary{calc,trees,positioning,arrows,chains,shapes.geometric,%
    decorations.pathreplacing,decorations.pathmorphing,shapes,%
    matrix,shapes.symbols}

\tikzstyle{line} = [draw, -, thick]
\tikzstyle{nodraw} = [draw, fill, circle, minimum width=0pt, inner sep=0pt]
\tikzstyle{box} = [line, rectangle, rounded corners, text centered]

\tikzset{
>=stealth',
  punktchain/.style={
    rectangle, 
    rounded corners, 
    draw=black, very thick,
    text width=2em, 
    minimum height=3em, 
    text centered, 
    on chain},
  line/.style={draw, thick, <-},
  element/.style={
    tape,
    top color=white,
    bottom color=blue!50!black!60!,
    minimum width=1em,
    draw=blue!40!black!90, very thick,
    text width=2em, 
    minimum height=3.5em, 
    text centered, 
    on chain},
  every join/.style={->, thick,shorten >=1pt},
  decoration={brace},
  tuborg/.style={decorate},
  tubnode/.style={midway, right=2pt},
}

\begin{document}
\title{Combining checkpointing and data compression to accelerate adjoint-based optimization problems}
\titlerunning{Accelerating adjoint checkpointing with compression}
%
\author{Navjot Kukreja\inst{1}\and
Jan H\"uckelheim\inst{1}\and Mathias Louboutin \inst{2} \and
Paul Hovland\inst{3} \and Gerard Gorman\inst{1}}
\authorrunning{N. Kukreja et al.}
%
\institute{Imperial College London, UK \email{nkukreja@imperial.ac.uk} \and
Georgia Institute of Technology, Atlanta, GA, USA
\\
\and
Argonne National Laboratory, Lemont, IL, USA\\
}
\maketitle              


\begin{abstract}
Seismic inversion and imaging are adjoint-based optimization problems that process up to terabytes of data, regularly exceeding the memory
capacity of available computers. Data compression is an effective strategy to
reduce this memory requirement by a certain factor, particularly if some loss in
accuracy is acceptable. A popular alternative is checkpointing, where data is
stored at selected points in time, and values at other times are recomputed as
needed from the last stored state.  This allows arbitrarily large adjoint
computations with limited memory, at the cost of additional recomputations.

In this paper, we combine compression and checkpointing for the first
time to compute a realistic seismic inversion. The combination of
checkpointing and compression allows
larger adjoint computations compared to using only compression, and
reduces the recomputation overhead significantly compared to using only checkpointing.
\end{abstract}

\section{Introduction}
\subsection{Adjoint-based optimization}
Adjoint-based optimization problems typically consist of a simulation
that is run forward in simulation time, producing data that is used in
reverse order by a subsequent adjoint computation that is run
backwards in simulation time. Many important
numerical problems in science and engineering use adjoints and follow
this pattern. 

Since the data for each of the computed timestep in the forward simulation will be used later in the adjoint computation, it would be
prudent to store it in memory until it is required again. However, the total size of this data
can often run into tens of terabytes, exceeding the memory capacity of most computer systems. Previous work has studied recomputation or data compression strategies to work around this problem. In this paper we investigate a combination of compression and recomputation.

\subsection{Example adjoint problem: Seismic inversion}
Seismic inversion typically involves the simulation of the propagation
of seismic waves through the earth's subsurface, followed by a
comparison with data from field measurements. The model of the
subsurface is iteratively improved by minimizing the misfit between
simulated data and field measurement in an adjoint optimization
problem~\cite{plessix2006review}. The data collected in an offshore survey typically consists of a
number of ``shots'' - each of these shots corresponding to different
locations of sources and receivers. Often  the gradient is computed for each of these
shots independently on a single cluster compute node, and then collated across all the shots to form a
single model update. The processing across shots is thereby easily parallelized and
requires only little communication, followed by a long
period of independent computation (typically around 10-100 minutes). Since the number of shots is typically of the order of
$10^4$, clusters can often be fully utilized even if individual shots are only processed on a single
node.

\subsection{Memory requirements}
A number of strategies have been studied to cope with the amount of data that occurs in adjoint computations - perhaps the simplest is to store all data to a
disk, to be read later by the adjoint pass in reverse order. However,
the computation often
takes much less time than the disk read and write, hence leaving disk speed as a bottleneck. 

Domain decomposition, where a single shot may be distributed across more
than one node, is often used not only to distribute the computational
workload across more processors, but also across more memory.. While this strategy is very powerful, the
number of compute nodes and therefore the amount of memory that can be used
efficiently is limited, for example by communication overheads that start to
dominate as the domain is split into increasingly small
pieces. Secondly, this strategy can be wasteful if the need for memory causes more nodes to be used than can be completely utilized for computation.
Lastly, this method is not well suited for
cloud-based setups since it can complicate
the setup and performance will suffer due to the slow inter-node communication.

Checkpointing is yet another strategy to reduce the memory
overhead. Only a subset of the timesteps during the forward pass is
stored. Other timesteps are recomputed when
needed by restarting the forward pass from the last available stored
state. We discuss this strategy in section~\ref{sec:revolve}. Previous
work has applied checkpointing to seismic imaging and inversion
problems~\cite{symes2007reverse, datta2018asynchronous}.
 An alternative is data compression, which is discussed in
section~\ref{sec:compression}. 

In this paper, we extend the previous studies by \emph{combining} checkpointing
and compression. This is obviously useful when the data does not fit in the
available memory even after compression, for example for very large adjoint
problems, or for problems where the required accuracy limits the achievable
compression ratios.

Compared to the use of only checkpointing without compression, this
combined method often improves performance. This is a consequence of
the reduced size of stored timesteps, allowing more timesteps to be
stored during the forward computation. This in turn reduces the amount
of recomputation that needs to be performed. On the other hand, the
compression and decompression itself takes time. The answer to the
question ``does compression pay off?'', depends on a number of factors including - available
memory, the required precision, the time taken to compress and
decompress, and the achieved compression factors, and various problem specific
parameters like computational intensity of the kernel involved in the
forward and adjoint computations, and the number of timesteps.

Hence, the answer to the compression question depends not only on the
problem one is solving (within seismic inversion, there are numerous
variations of the wave equation that may be solved), but also the
hardware specifics of the machine on which it is being solved. In
fact, as we will see in section \ref{sec:errors}, the answer might even change
during the solution process of an individual problem. This brings up
the need to predict whether compression pays off in a
given scenario, without incurring significant overheads in answering
this question. To this end, we present a performance
model that answers that question.

\subsection{Summary of contributions}
In this paper, we study
\begin{itemize}
\item the use of different compression algorithms to seismic
  data including six lossless and a lossy compression
  algorithm for floating point data,
\item a performance model for checkpointing alone, taking into
  account the time taken to read and write checkpoints, and
\item an online performance model to predict whether compression would speed up
  an optimization problem.
\end{itemize}

\section{Compression algorithms}

\label{sec:compression}

Data compression is increasingly used to reduce the memory footprint of
scientific applications. This has been accelerated by the advent of special purpose compression
algorithms for floating-point scientific data, such as ZFP
or SZ~\cite{lindstrom2014fixed,di2018efficient}.

Lossless algorithms guarantee that the exact original data can be recovered
during decompression, whereas lossy algorithms introduce an error, but often
guarantee that the error does not exceed certain absolute or relative error
metrics. Typically, lossy compression is more effective in reducing the data
size. Most popular compression packages offer various settings that allow a
tradeoff between compression ratio, accuracy, and compression and decompression
time.

Another difference we observed between lossless and lossy
compression algorithms was that the lossless compression algorithms we evaluated tended to
interpret all data as one-dimensional series only while SZ and ZFP,
being designed for scientific data, take the dimensionality
into account directly. This makes a difference in the case of a
wavefield, for example, where the data to be compressed corresponds to
a smoothly varying function in (two or) three dimensions and
interpreting this three-dimensional data as one-dimensional would
completely miss the smoothness and predictability of the data values.

It is worth noting that another data reduction strategy is to typecast values
into a lower precision format, for example, from double precision to single
precision. This can be seen as a computationally cheap lossy compression
algorithm with a compression ratio of $2$.

Perhaps counterintuitively, compression can not only reduce the memory
footprint, but also speed up an application. Previous work has observed that the
compression and decompression time can be less than the time saved from the
reduction in data that needs to be communicated across MPI nodes or between a
GPU and a host computer~\cite{gpu-compression}.

One way of using compression in adjoint-based methods is to compress
all timesteps during the forward pass. If the compression ratio is
sufficient to fit the compressed data in memory, compression can
serve as an \emph{alternate strategy} to checkpointing. Previous work has discussed this in the context of
computational fluid dynamics~\cite{cyr2015towards,marin2016large} and seismic
inversion using compression algorithms specifically designed for
the respective applications~\cite{dalmau2014lossy,boehm2016wavefield}. 

Since the time spent on compressing and decompressing data is often
non-negligible, this raises the question whether the computational
time is better spent on this compression and decompression, or on the
recomputation involved in the more traditional checkpointing
approach. This question was previously answered to a limited extent
for the above scenario where compression is an alternative to
checkpointing, in a specific application~\cite{cyr2015towards}. We discuss
this in more detail in section \ref{sec:perf_compression}. 

\subsection{Lossless Compression}
Blosc is a library that provides optimized high-performance implementations of various lossless compressors, sometimes beyond their corresponding reference implementations~\cite{alted2010modern}. For our experiments we use this library through its python interface. The library includes implementations for
six different lossless compression algorithms, namely ZLIB, ZSTD, BLOSCLZ,
LZ4, LZ4HC and Snappy. All these algorithms look at the data as a one-dimensional stream of bits
and at least the blosc implementations have a limit on the size of the one-dimensional array that
can be compressed in one call. Therefore we use the python package \emph{blosc-pack}, which is
a wrapper over the blosc library, to implement \emph{chunking}, i.e. breaking up the stream into
chunks of a chosen size, which are compressed one at a time. 

\subsection{Lossy Compression}
We use the lossy compression package ZFP~\cite{lindstrom2014fixed} written in
C. To use ZFP from python, we developed a python wrapper for the reference
implementation of ZFP \footnote{To be released open source on publication}. ZFP supports three compression modes, namely fixed tolerance, fixed precision
and fixed rate. The fixed-tolerance mode limits the absolute error, while the
fixed-precision mode limits the error as a ratio of the range of values in the array to be compressed.
The fixed-rate mode achieves a guaranteed compression ratio requested by the
user, but does not provide any bounds on accuracy loss.

The fixed-rate mode could make our implementation more straightforward by offering a predictable size of compressed checkpoints, but the lack of error bounds makes this option less attractive. Moreover, ZFP claims to achieve the best "compression efficiency" in the fixed-tolerance mode, and we thus chose to focus on this mode.

SZ~\cite{di2018efficient} is a more recently developed compression library, also focussed on lossy compression
of floating-point scientific data, also developed in C. While we have also written a python wrapper for the reference
implementation of SZ \footnote{Also to be released open source
upon publication}, a thorough comparison of ZFP and SZ remains future work.



\section{Checkpointing performance model}
\label{sec:revolve}
As previously mentioned, checkpointing is a strategy to store selected timesteps, and recompute others when needed. The question which checkpoints should be stored to get the best tradeoff between recomputation time and memory footprint was answered in a provably optimal way by the Revolve checkpointing algorithm~\cite{griewank2000algorithm}. Revolve makes certain assumptions, for example that all timesteps have the same compute cost and storage size, the number of timesteps is known a priori, and there is only one level of memory (e.g. RAM) that is restricted in size, but very fast. Other authors have subsequently developed
extensions to Revolve that are optimal under different
conditions~\cite{stumm2009multistage,
aupy2016optimal}. We focus in this paper on the classic Revolve algorithm, and store all checkpoints in RAM.

In this section, we build on the ideas introduced in \cite{stumm2009multistage} to build a performance model that predicts the runtime of an adjoint computation using Revolve checkpointing. 
We call the time taken by a single forward computational step $C_F$
and correspondingly, the time taken by a single backward step $C_R$. For a simulation with $\mathbf{N}$ timesteps, the minimum wall time required
for the full forward-adjoint evaluation is given by
\begin{equation}
T_N = \mathbf{C_F} \cdot \mathbf{N} + \mathbf{C_R} \cdot \mathbf{N}
\end{equation}
If the size of a single timestep in memory is given by $\mathbf{S}$, this
requires a memory of at least size $\mathbf{S} \cdot \mathbf{N}$. If sufficient memory
is available, no checkpointing or compression is needed.

If the memory is smaller than $\mathbf{S} \cdot \mathbf{N}$, Revolve provides
a strategy to solve for the adjoint field by storing a subset of the $\mathbf{N}$ total checkpoints
and recompute the remaining ones. The overhead introduced by this method can be broken down into
the recomputation overhead $\mathbf{O}_R$ and the storage overhead $\mathbf{O}_S$. The recomputation
overhead is the amount of time spent in recomputation, given by
\begin{equation}
\mathbf{O}_R(N, M) = p(N, M) \cdot \mathbf{C_F},
\end{equation}
where $p(N, M)$ is the minimum number of recomputed steps from \cite{griewank2000algorithm}, given as
\begin{equation}
p(N, M) = \begin{cases}
      N(N-1) /2, & \text{if}\ M=1 \\
      \min\limits_{1<=\widetilde{N}<=N} \{\widetilde{N} + p(\widetilde{N}, M) + p(N-\widetilde{N}, M-1)\}, & \text{if}\ M>1
    \end{cases}
    \label{eqn:recompute}
\end{equation}
where $M$ is the number of checkpoints that can be
stored in memory. Note that for $M \geq N$, $\mathbf{O}_R$ would be zero. For $M <
N$, $\mathbf{O}_R$ grows rapidly as M is reduced relative to N. 

In an ideal implementation, the storage overhead $\mathbf{O}_S$ might be zero, since the computation could
be done ``in-place'', but in practice, checkpoints are generally stored in a separate section of memory and they
need to be transferred to a ``computational'' section of the memory where the computation is performed, and then
the results copied back to the checkpointing memory. This copying is a common feature of checkpointing
implementations, and might pose a non-trivial overhead when the
computation involved in a single timestep is not very large. 
This storage overhead is given by:
\begin{equation}
\mathbf{O}_{SR}(N, M) = \mathbf{W}(N, M) \cdot \frac{\mathbf{S}}{\mathbf{B}} +
\mathbf{N} \cdot \frac{\mathbf{S}}{\mathbf{B}}
\label{eqn:storage}
\end{equation}
where $\mathbf{W}$ is the total number of times Revolve writes
checkpoints for a single run, $ \mathbf{N}$ is the number of times
checkpoints are read, and $\mathbf{B}$ is the bandwidth at which these
copies happen. The total time to solution becomes
\begin{equation}
T_R = \mathbf{C_F} \cdot \mathbf{N} + \mathbf{C_R} \cdot \mathbf{N} + \mathbf{O}_R(N, M) +
\mathbf{O}_{SR}(N, M)
\end{equation} 

\section{Performance model including compression}
\label{sec:perf_compression}
By using compression, the size of each checkpoint is reduced and the
number of checkpoints available is increased ($M$ in equation
\ref{eqn:recompute}). This reduces the recomputation overhead $\mathbf{O}_R$,
while at the same time adding overheads related to compression and decompression
in $\mathbf{O}_S$.
To be beneficial, the reduction in $\mathbf{O}_R$ must offset the increase in 
$\mathbf{O}_{SR}$, leading to an overall decrease in the time to solution $T$.

Our performance model assumes
 that the compression algorithm behaves uniformly
across the different time steps of the simulation, i.e. that we get the same compression ratio, compression time and 
decompression time, no matter which of the $N$ possible checkpoints we try to compress/decompress. The storage overhead
now becomes
\begin{equation}
\begin{split}
\mathbf{O}_{SR}(N, M) = &\mathbf{W}(N, M \cdot F) \cdot \left(\frac{\mathbf{S}}{\mathbf{F}
  \cdot \mathbf{B}} + t_c\right) +\\&\mathbf{N} \cdot
\left(\frac{\mathbf{S}}{\mathbf{F} \cdot \mathbf{B}} + t_d\right)
\end{split}
\end{equation}
where $\mathbf{F}$ is the compression ratio (i.e. the ratio between the uncompressed and compressed checkpoint), and $t_c$
and $t_d$ are compression and decompression times, respectively. At the same
time, the recomputation overhead decreases
because $\mathbf{F}$ times more checkpoints are now available.

\section{Acceptable errors and convergence}
\label{sec:errors}

Our performance model is agnostic of the specific 
optimization problem being solved. We envision it being used in a generic
checkpointing runtime that manages the checkpointed execution of an optimization
problem, and accepts an acceptable error tolerance as an input parameter for each
gradient evaluation and determines whether or not compression can pay off for that
iteration. For this reason, we do not discuss in this paper whether or not a certain accuracy is acceptable for any given application.

We note that there is some previous work in this area, discussing for example  the effect of bounded pointwise errors in a multi-dimensional field on computed numerical derivatives, for ZFP \cite{zfp-derivatives}. In the context of seismic inversion, other work discusses accuracy requirements in optimization loops, and notes that high accuracy is only needed when already close to a minimum \cite{van20143d, boehm2016wavefield}. There has been previous work on choosing the most appropriate compression algorithm under some circumstances
\cite{tao2018optimizing}., and work that addresses convergence guarantees of trust-region
based optimization methods in the presence of gradients that are only known
with a probability $p$.~\cite{blanchet2016convergence}

Despite all this previous work, for most practical adjoint optimization applications, the relationship between accuracy (whether caused by roundoff, compression or truncation errors) and convergence remains a field of ongoing research.

\section{Problem and test case}
\label{sec:devito}
We use Devito~\cite{devito-api} to solve forward and adjoint
wave equation problems. Devito is a domain-specific language that enables the
rapid development of finite-difference solvers from a high-level description
of partial differential equations. The simplest version of the seismic wave
equation is the acoustic isotropic wave equation defined as:
\begin{equation}
m(x)\frac{\partial^2 u(t, x)}{\partial t^2} - \nabla^2 u(t, x) = q(t, x),
\label{eqn:wave}
\end{equation}
where $m(x) = \frac{1}{c^2(x)}$ is the squared slowness, $c(x)$ the spatially
dependent speed of sound, $u(t, x)$ is the pressure wavefield, $\nabla^2 u(t, x)$ denotes the laplacian of the wavefield and $q(t, x)$ is a source term.

The solution to equation~\ref{eqn:wave} forms the forward problem. The seismic inversion problem minimizes the misfit between simulated and observed signal given by:
\begin{equation}
\min_{m} \phi_s(m) = \frac{1}{2} \left\lVert d_{sim} - d_{obs} \right\rVert_2^2.
\end{equation}

We call the kernel derived from a basic finite difference formulation of Equation~\ref{eqn:wave}, the OT2 kernel because it is second-order accurate in time. We also use another formulation from \cite{devito-api}, which is 4th-order accurate in time. We call this the OT4 kernel.

This optimization problem is usually solved using gradient based methods such as steepest descent,
where the gradient is computed using the adjoint-state method.

The values of $m(x)$ used in this work are derived from the Overthrust
model~\cite{aminzadeh1996three} over a grid of $287 \times 881 \times
881$ points, including an absorbing layer of 40 points on each
side. The grid spacing is $25m$ in space. The propagation time is
$4sec$ that corresponds to 2526 timesteps. The wave field at the final
time is shown in Figure~\ref{fig:uncompressed}. The uncompressed size
of this single time step field is just under 900MB. If one were to
store all the timesteps, this would require 2.3TB of memory.

To implement Revolve with Devito, we use pyRevolve~\cite{kukreja2018high} which
is a python library to manage the execution of checkpointed adjoint computations. The performance model in
section~\ref{sec:revolve} assumes that the implementation is similar
to pyRevolve, which stores a checkpoint by copying a portion of the operator's
working memory to the checkpointing memory and similarly loads a checkpoint by
copying from the checkpointing memory to the operator's working memory.

For benchmarking we used a dual-socket
Intel(R) Xeon(R) Platinum 8180M @ 2.50 Ghz (28 cores each) (skylake).

\section{Results and discussion}
\label{sec:results}
\label{sec:compressibility_along_time}
\begin{figure}[htb]
\begin{center}
\includegraphics[width=0.8\linewidth]{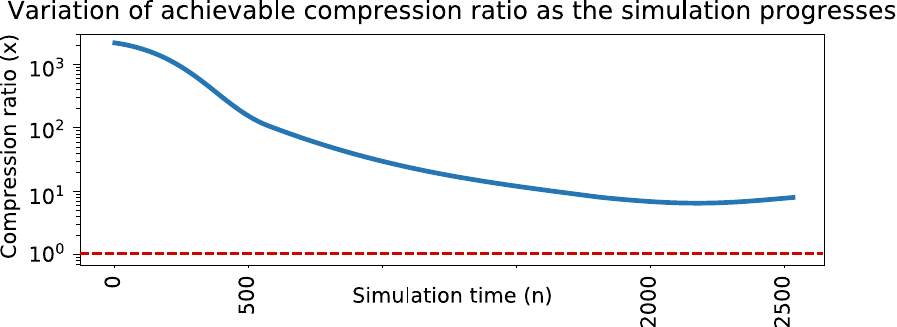}
\end{center}
\caption{Compression ratios achieved on compressing different time steps. Every timestep from 1 to 2526 was compressed and plotted.}
\label{fig:compression_ratios_through_simulation}
\end{figure}
To understand the compressibility of the data produced in a typical
wave-propagation simulation, we ran a simulation as per the setup
described in section \ref{sec:devito}, and tried to compress every single timestep. For this we chose ZFP in fixed tolerance mode at some arbitrary tolerance level. We noted the compression ratios achieved at every
timestep. As figure \ref{fig:compression_ratios_through_simulation} shows, the initial timesteps are much easier
to compress than the later ones. This is not surprising since most
wave simulations start with the field at rest, i.e. filled with zeros. As the wave reaches more parts of the domain, the field becomes less
compressible until it achieves a stable state when the wave has
reached most of the domain. 

If the simulation had started with the field already oscillating in a
wave, it is likely that the compressibility curve for that simulation
would be flat. This tells us that the compressibility of the last timestep of the
solution is representative of the worst-case compressibility and hence
we used the last timestep as our reference for comparison of
compression in the rest of the analysis.

\begin{table*}[htb]
\resizebox{\textwidth}{!}{
\begin{tabular}[c]{| c | c | c | c | c | c | c |}
\hline
 Compressor & Chunk size(bytes) & Shuffle Mode & Setting & Compression time(ms) & Decompression time(ms) & Compression Ratio \\
 \hline
 BloscLZ & 1048576 & SHUFFLE & 6 & 4249.44 & 1288.86 & 1.188 \\
 LZ4 & 2965280 & SHUFFLE & 4 & 1371.26 & 920.98 & 1.199 \\
 LZ4HC & 2097152 & SHUFFLE & 8 & 31245.16 & 926.69 & 1.265\\
 ZLib & 524288 & SHUFFLE & 7 & 30218.81 & 2470.04 & 1.291\\
 ZStd & 524288 & SHUFFLE & 9 & 117238.76 & 1477.34 & 1.312\\
\hline
\end{tabular}
}
\caption{Some results from trying out all possible compressors and settings in blosc. We selected the best compression ratio seen for each compressor. "Setting" here is the choice between speed and compression, where 0 is fastest and 9 is highest compression.}
 \label{tbl:lossles}
\end{table*}
Table \ref{tbl:lossles} shows the compression ratios and times for a few different lossless compressors and their corresponding settings. As can be seen, the compression factors achieved, and the time taken to compress and decompress can vary significantly, but it is hard to say whether this compression could be used to speed up the inversion problem.

\begin{figure}[htb]
\begin{subfigure}[b]{0.5\textwidth}

\includegraphics[width=\linewidth]{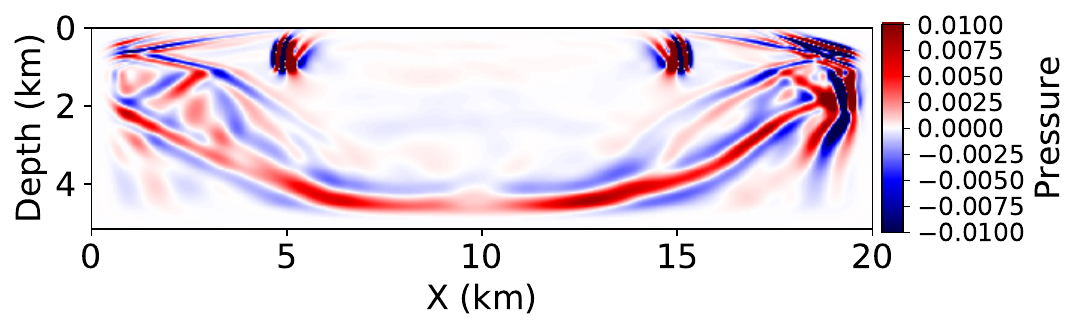}

\caption{Reference wavefield for
  compression and decompression. }
\label{fig:uncompressed}
\end{subfigure}
\begin{subfigure}[b]{0.5\textwidth}
\includegraphics[width=\linewidth]{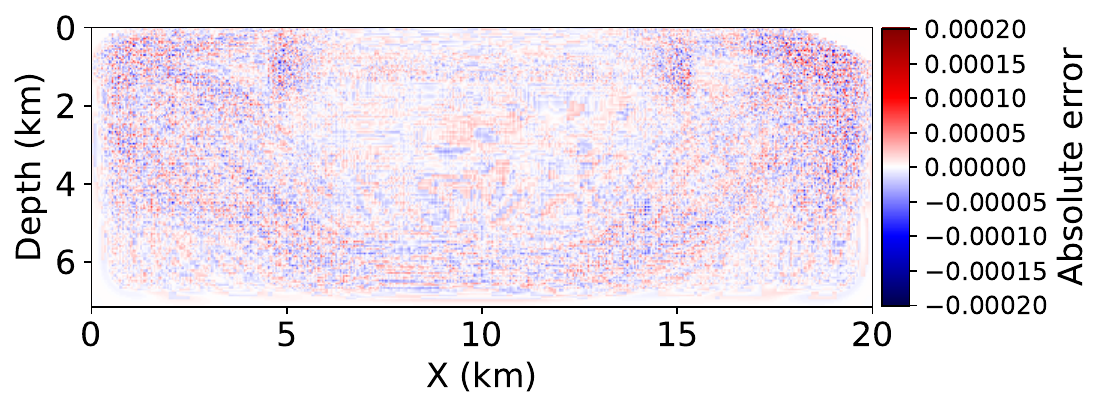}

\caption{Errors introduced
  during compression and decompression using the fixed-tolerance
  mode.}
\label{fig:decompressed_error}
\end{subfigure}
\caption{This field was formed after a Ricker
  wavelet source was placed at the surface of the model and the wave propagated for 2500
  timesteps. This is a vertical (x-z) cross-section of a 3D field, taken at
  the $y$ source location.  It is interesting to note that the errors are more or less
  evenly distributed across the domain with only slight variations
  corresponding to the wave amplitude (from Figure
  \ref{fig:uncompressed}). A small block-like structure characteristic of
  ZFP can be seen.}
\end{figure}

\begin{figure}[htb]
\begin{subfigure}{0.5\linewidth}
\includegraphics[width=\textwidth]{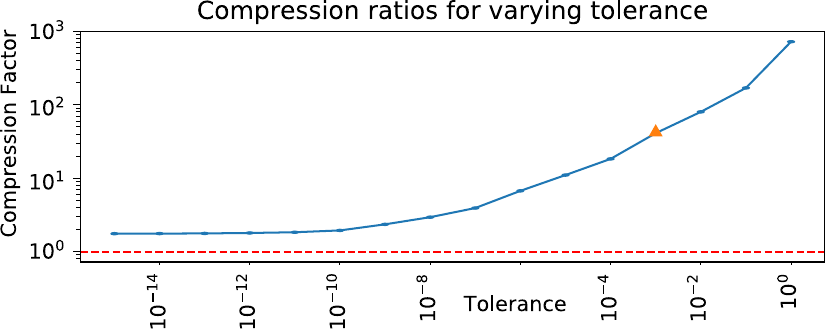}
\caption{Effect of tolerance on Compression Ratio}
\label{fig:tolerance_cf_plot}

\end{subfigure}
\begin{subtable}{0.5\linewidth}
    \centering
\begin{tabular}{|c|c|}
\hline
     Tolerance & Gradient error \\
     \hline
     0.1 & 662.905 \\
     0.01 & 70.619 \\
     0.001 & 10.485 \\
     0.0001 & 0.763 \\
     $10^{-5}$ & 0.194 \\
     $10^{-6}$ & 0.154 \\
     $10^{-7}$ & 0.151 \\
     \hline
     
\end{tabular}
    \caption{Effect of tolerance on Gradient error}
    \label{tbl:gradient_errors}

\end{subtable}
\caption{Effect of tolerance settings of ZFP in fixed-tolerance mode on Compression ratio (left) and final gradient evaluation (right). We
  define compression ratio as the ratio between the size of the
  uncompressed data and the compressed data. The dashed line
  represents no compression. The highlighted point corresponds to the
  setting used for the other results here unless otherwise specified. The gradient error (right) is the 2-norm of the error tensor in the gradient, as compared with an exact computation.}
\end{figure}

Figure~\ref{fig:tolerance_cf_plot} shows compression ratios for different tolerance 
settings for the fixed-tolerance mode of ZFP. The point highlighted here was the setting used to compress all timesteps in figure \ref{fig:compression_ratios_through_simulation}. Figure
\ref{fig:decompressed_error} shows the spatial distribution of the
errors after compression and decompression, compared to the original
field, for this setting. Table \ref{tbl:gradient_errors} shows the effect of different levels of pointwise absolute error on the overall error in the gradient evaluation. We can see that the error in the gradient evaluation does not explode. 

\begin{figure}[htb]
    \centering
    \begin{subfigure}[b]{0.32\textwidth}
    \includegraphics[width=\textwidth]{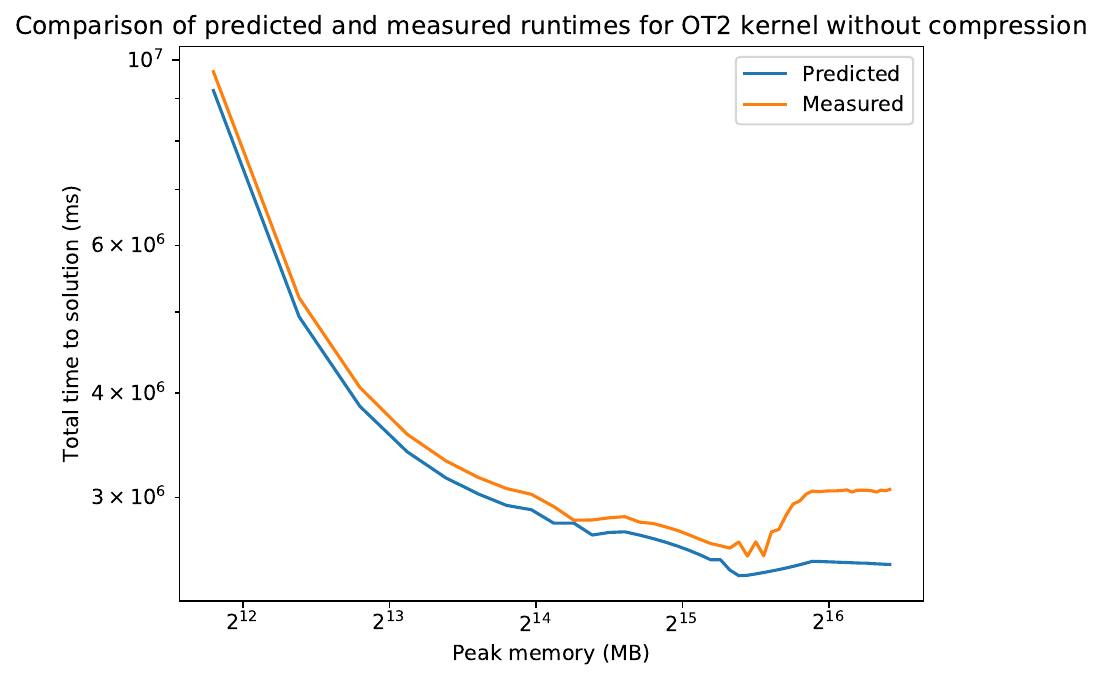}
    \caption{OT2, No compression}
    \label{fig:validation_nocompression}
    \end{subfigure}
    \begin{subfigure}[b]{0.32\textwidth}
    \includegraphics[width=\textwidth]{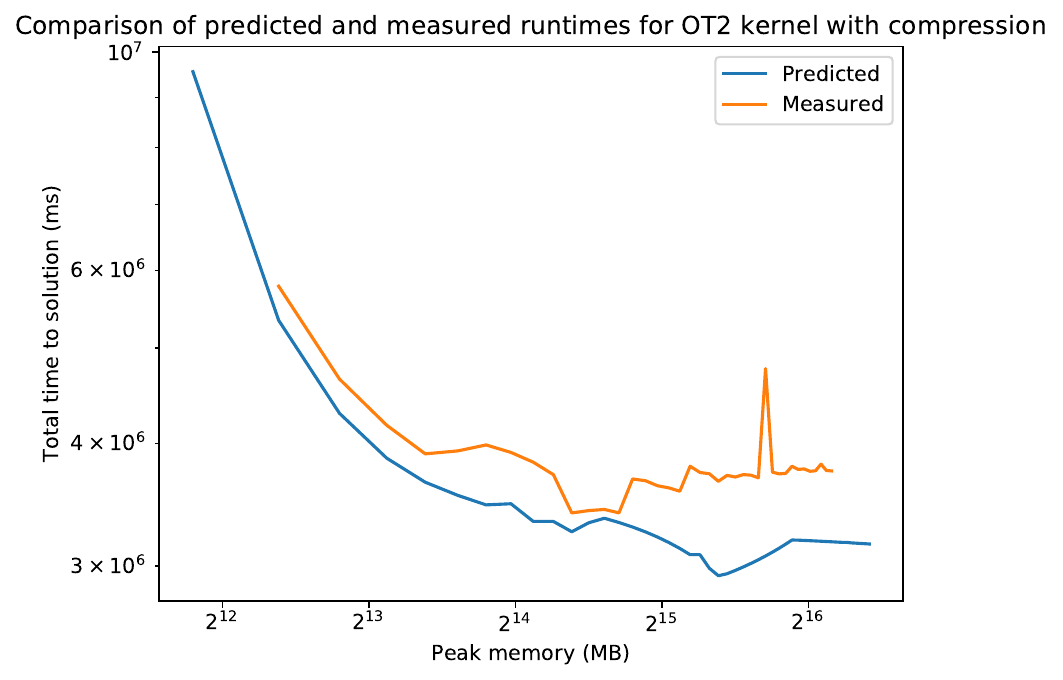}
    \caption{OT2, Compression}
    \label{fig:validation_compression_ot2}
    \end{subfigure}
    \begin{subfigure}[b]{0.32\textwidth}
    \includegraphics[width=\textwidth]{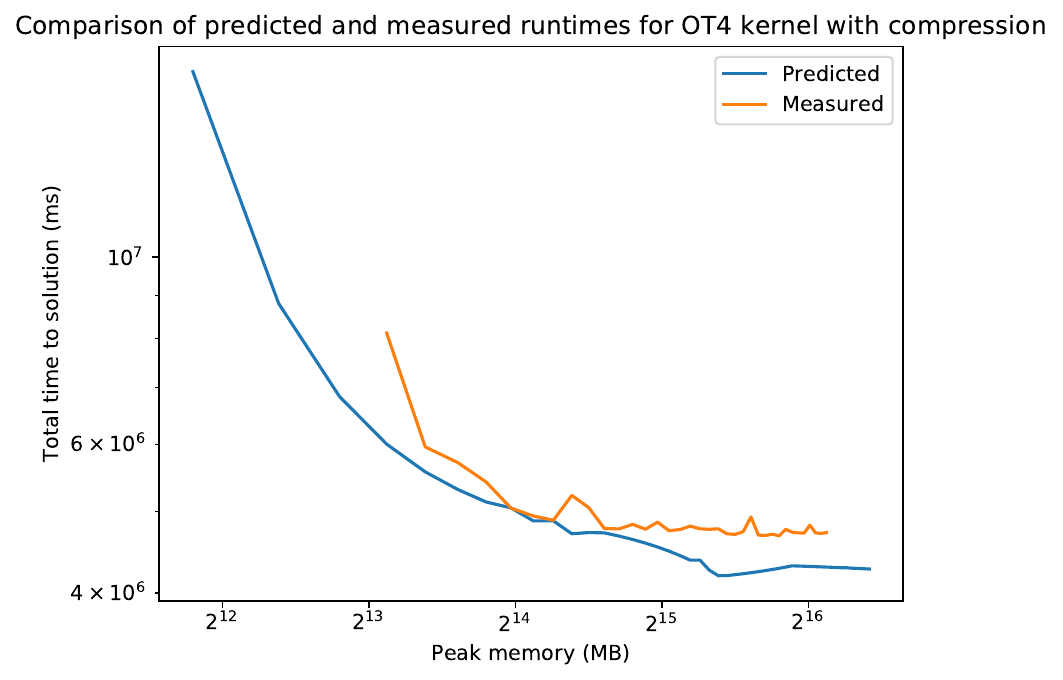}
    \caption{OT4, Compression}
    \label{fig:validation_compression_ot4}
    \end{subfigure}
    \caption{Predicted vs measured runtimes for two different kernels (OT2 and OT4), with and without compression. This shows that the performance model can predict the runtime effectively. The compression setting used was ZFP with absolute error tolerance set to $10^{-6}$}
\end{figure}
To validate the revolve-only performance model, figure \ref{fig:validation_nocompression} shows the predicted runtime for a variety of peak memory constraints along with measured runtime for the same scenario. Figure \ref{fig:validation_compression_ot2} shows a comparison of predicted and measured runtimes for the OT2 kernel with compression enabled. Figure \ref{fig:validation_compression_ot4} repeats this experiment for the OT4 kernel which has a higher computational complexity. 
It can be seen that the model is able to predict the real performance very closely in all three cases.

\begin{figure}[htb]
\begin{center}
\vspace{-1em}
\includegraphics[width=0.6\linewidth]{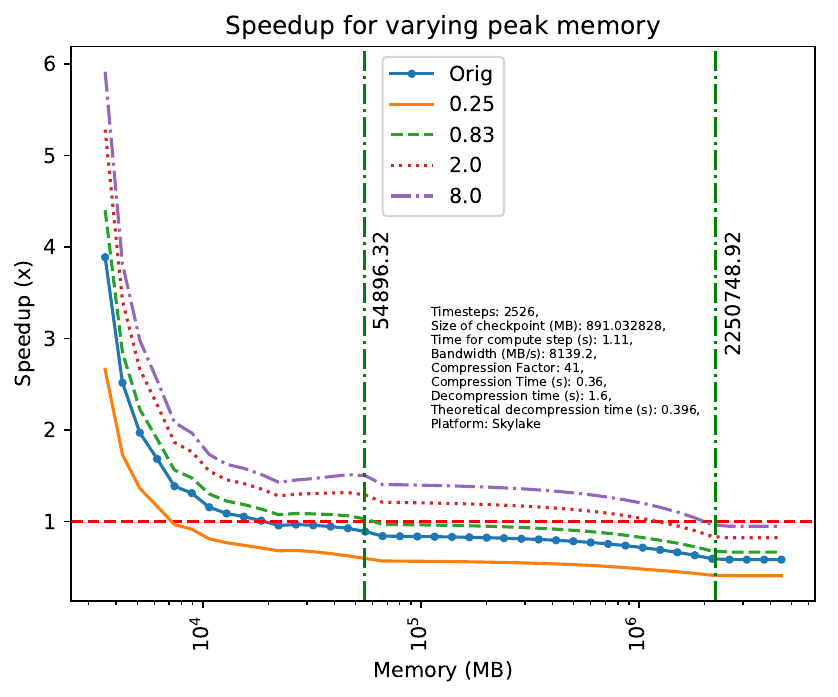}
\vspace{-1em}
\end{center}
\caption{The speedups predicted by the performance model for varying
  memory. The baseline
(1.0) is the performance of a Revolve-only implementation under the
same conditions. The different curves represent kernels with differing
compute times (represented here as a factor of the sum of compression
and decompression times). The first vertical line at $~$ 53GB marks the
spot where the compressed wavefield can completely fit in memory and
Revolve is unnecessary if using compression. The second vertical line
at $~$ 2.2 TB marks the spot where the entire uncompressed wavefield can
fit in memory and neither Revolve nor compression is necessary. The
region to the right is where these optimizations are not necessary or
relevant. The middle region has been the subject of past studies using
compression in adjoint problems. The region to the left is the focus
of this paper.}
\label{fig:varying_memory}
\end{figure}

We have now seen that the performance model from Section~\ref{sec:perf_compression} is effective at predicting the runtime of adjoint computations. 
To study the performance model, we first visualize it along the axis of available memory, comparing the predicted performance of the chosen compression scheme with the predicted performance of a Revolve-only adjoint implementation. This is shown in Figure~\ref{fig:varying_memory} where we can distinguish three different scenarios, depending on the amount of
available memory.
\begin{enumerate}
\item If the memory is insufficient even with compression to store the entire
trajectory, one can either use checkpointing only, or combine checkpointing with compression. This is the left section of the figure.
\item If the available memory is not sufficient to store the uncompressed
trajectory, but large enough to store the entire compressed trajectory, we compare two possible strategies: Either use compression only, or use checkpointing only. This is the middle section of the figure.
\item If the available system memory is large enough to hold the entire
uncompressed trajectory, neither compression nor checkpointing is necessary. This is the right section of the figure.
\end{enumerate}

The second scenario was studied in previous work~\cite{cyr2015towards}, while
the combined method is also applicable to the first scenario, for which previous work has only used checkpointing without compression.

\begin{figure}[htb]
\begin{subfigure}[b]{0.5\textwidth}
\begin{center}
\includegraphics[width=\linewidth]{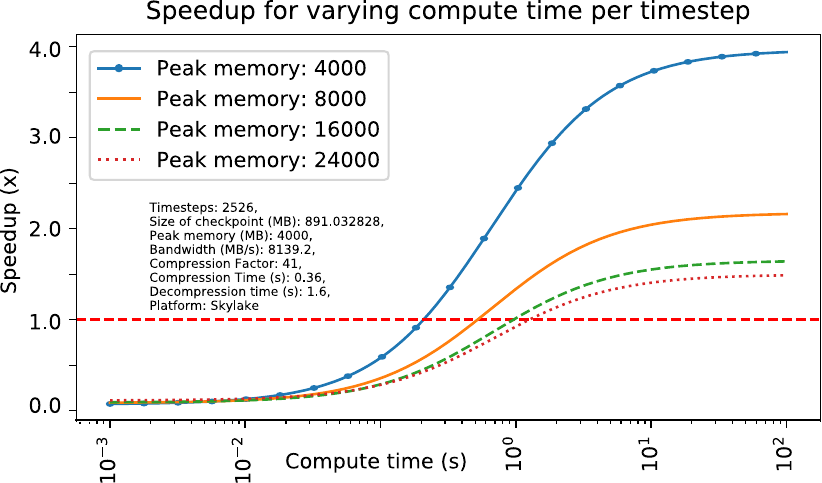}
\end{center}
\caption{Varying Compute}
\label{fig:varying_compute}
\end{subfigure}
\begin{subfigure}[b]{0.5\textwidth}
\begin{center}
\includegraphics[width=\linewidth]{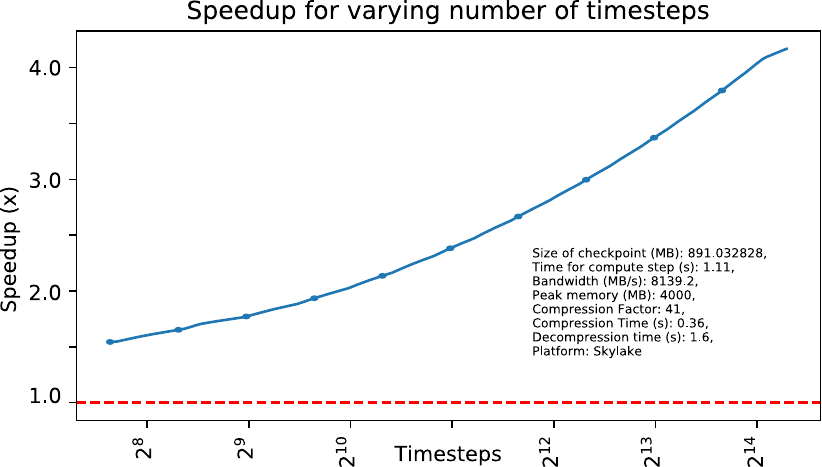}
\end{center}
\caption{Varying timesteps}
\label{fig:varying_nt}

\end{subfigure}
\caption{The speedups predicted by the performance model for varying
  compute cost (left) and number of timesteps (right). The baseline
(1.0) is the performance of a Revolve-only implementation under the
same conditions. The benefits of compression drop rapidly if the
computational cost of the kernel that generated the data is much lower
than the cost of compressing the data. For increasing computational
costs, the benefits are bounded. It can be seen that compression becomes more
beneficial as the number of timesteps is increased.}
\end{figure}

We can identify a number of factors that make compression more likely to be
beneficial compared to pure checkpointing: A very small system memory size and a
large number of time steps lead to a rapidly increasing recompute factor, and
compression can substantially reduce this recompute factor. This can be seen in
Figures~\ref{fig:varying_memory} and~\ref{fig:varying_nt}.

The extent to which the recompute factor affects the overall runtime also
depends on the cost to compute each individual time step. If the compute cost
per time step is large compared to the compression and decompression cost, then
compression is also likely to be beneficial, as shown in
Figure~\ref{fig:varying_compute}. As the time per time step increases and the
compression cost becomes negligible, we observe that the ratio between the runtime
of the combined method and that of pure checkpointing is only determined by the
difference in recompute factors.

\section{Conclusions and Future work}
We used  compression to reduce the computational overhead of
checkpointing in an adjoint computation used in seismic
inversion. We developed a performance model
that computes whether or not the combination of compression and
checkpointing will outperform pure checkpointing or pure compression
in a variety of scenarios, depending on the available memory size,
computational intensity of the application, and compression ratio and
throughput of the compression algorithm. In future work, we plan to extend this work by
\begin{itemize}
\item further exploring the relationship between pointwise error bounds in compression and the overall error of the adjoint gradient evaluation,
\item extending our performance model to support non-uniform compression ratios, as would be expected for example if the initial wave field is smoother and therefore mnore easily compressible,
\item studying strategies where different compression settings (or even no compression) is used for a subset of time steps,
\item exploring compression and multi-level checkpointing, including SSD or hard drives in addition to RAM storage,
\item and finally by developing checkpointing strategies that are optimal even if the size of checkpoints post-compression varies and is not known a priori.
\end{itemize}

\section*{Acknowledgments}
This work was funded by the Intel Parallel Computing Centre at
Imperial College London and EPSRC EP/R029423/1. 
This work was supported by the U.S. Department of Energy, Office of Science,
Office of Advanced Scientific Computing Research, Applied Mathematics and
Computer Science programs under contract number DE-AC02-06CH11357.
We would also like to acknowledge the support from the SINBAD II project and
the member organizations of the SINBAD Consortium.

We gratefully acknowledge the computing resources provided and operated by the
Joint Laboratory for System Evaluation (JLSE) at Argonne National Laboratory.

This paper benefitted from discussions with Kaiyuan Huo, Fabio Luporini, Thomas Matthews, Paul Kelly, Oana Marin.

\small
\bibliographystyle{splncs04}
\bibliography{compression}


\end{document}